\documentclass[prd,aps,groupedaddress,showpacs]{revtex4}
\begin{document}
%\draft

\title{Fluctuations of Quantum Radiation Pressure in Dissipative Fluid}

\author{Chun-Hsien Wu}

\email{chunwu@phys.sinica.edu.tw}

\author{Da-Shin Lee}

\email{dslee@mail.ndhu.edu.tw}

\affiliation{Department of Physics, National Dong-Hwa University,
Hualien, Taiwan, R.O.C.}

\vskip 2cm

\begin{abstract}
Using the generalized Langevin equations involving the stress
tensor approach, we study the dynamics of  a perfectly reflecting
mirror which is exposed to the electromagnetic radiation pressure
by a laser beam in a fluid at finite temperature.  Based on the
fluctuation-dissipation theorem, the minimum uncertainty of the
mirror's position measurement from both quantum and thermal noises
effects including the photon counting error  in the laser
interferometer is obtained in the small time limit as compared
with the "standard quantum limit".
 The result of the large time behavior of fluctuations of the
 mirror's velocity in a  dissipative environment  can be  applied to the laser
interferometer of the ground-based gravitational wave detector.

\end{abstract}

\date{\today{}}


\maketitle

It is known that quantum fluctuations of radiation pressure play
an essential role in limiting the sensitivity of very precise
quantum optical measurements in a laser interferometer which
measures small changes in the position of the end mirror for
detecting gravitational waves~\cite{Caves1,Caves2}. When a mirror
is exposed to a laser beam, its position is coupled to the laser
intensity via quantum radiation pressure. The exerted force due to
radiation pressure on the mirror can be expressed as the integral
of the expectation values of the Maxwell stress tensor operators
with respect to some physically realizable quantum states. While
the quantum states are not the eigenstates of the stress tensor
operators, the radiation force exhibits fluctuations. In
\cite{WF}, Wu and Ford have studied the effects of quantum
fluctuations of radiation pressure on a mirror. They obtained the
fluctuations in velocity and position of the mirror in terms of
the mean squared fluctuations of the stress tensor of quantum
electromagnetic fields that are known to suffer from a
state-dependent divergence in the coincidence limit~\cite{WF99}.
However, the integrals of these stress tensor correlation
functions over space and time that can be directly linked to the
mirror's velocity fluctuations are shown to be finite using the
method of integration by parts by assuming a switch-on-switch-off
procedure in the laser beam. In this Letter, we would like to
generalize this approach by investigating the motion of a mirror
driven by a fluctuating electromagnetic radiation force in a
dissipative fluid at finite temperature. By calculating the
velocity fluctuations of the mirror in this system, we can study
how the dynamics of quantum fluctuations of radiation pressure is
affected by the dissipative effects.

In classical statistical mechanics, the non-equilibrium dynamics
of the Brownian motion in a stationary fluid can be described by a
phenomenological but very successful approach, namely  the
Langevin equations. Incessant collisions of the molecules of the
fluid with the Brownian body produce both resistance to the motion
of the body as well as fluctuations in its trajectory. The
Langevin equations phenomenologically account for these effects by
introducing a term proportional to the velocity of the body that
incorporates friction and dissipation as well as a stochastic
force term that reflects the random kicks of the molecules in the
fluid on the body that are thus related by the
fluctuation-dissipation theorem . In most of applications, the
stochastic noise is assumed to be completely uncorrelated and the
coefficient of the friction term determines the relaxation time of
the body. Notice that these Langevin equations can  be
straightforwardly generalized by involving the stress tensor
approach to take account of the effects of the  fluctuating
quantum force due to  radiation pressure. In this work, we will
adopt this generalized Langevin equations to tackle the issue
under consideration.

Here we consider a mirror of mass $m$ which is oriented
perpendicularly to the $x$-direction and is exposed to the
electromagnetic radiation pressure in a dissipative fluid of
temperature $T$. We assume that the incident radiation by a laser
beam with a circular spot of radius $R$ exerts a fluctuating force
$F$ on the one side of the mirror in the $x$ direction within time
interval $\tau $. Then the corresponding Langevin equations to
describe the motion of the mirror which can be treated classically
in such a fluctuating environment is written as \cite{PRS,KJ}
\begin{equation}
m\frac{dv}{dt}=-\xi \, v+\eta (t)+F(t)\, ,\label{langevin}\end{equation}
 where $\xi $ is the friction coefficient and $\eta (t)$ is the
stochastic force to account for the random kicks of the molecules
of the fluid on the mirror. Its statistical properties can be summarized
as follows: \begin{eqnarray}
\langle \eta (t)\rangle  & = & 0\,  ; \nonumber \\
\langle \eta (t)\eta (t')\rangle  & = & 2\xi K_{\rm{B}}T\delta
(t-t')\, , \label{correlation}
\end{eqnarray}
 where $K_{\rm{B}}$ is Boltzmann's constant and the average is taken with respect to the ensemble of
 thermal equilibrium
fluctuations of the fluid at finite temperature. The radiation force on
the mirror in the $x$ direction can be expressed by the area
integral of the stress tensor \begin{equation} F=\int _{A}T_{xx}da
\, ,\end{equation}
 where $T_{ij}$ is the Maxwell stress tensor. When we treat the incident
radiation quantum-mechanically, the expectation value of the
stress tensor operator with respect to the quantum state of
radiation fields
is divergent due to vacuum fluctuations. We can renormalize it by
replacing the stress tensor operator $T_{ij}$ with a normal
ordered one $:T_{ij}:\,=T_{ij}-\langle T_{ij}\rangle _{0}$, which
means to subtract out the vacuum divergence $\langle T_{ij}\rangle
_{0}$ such that $\langle :T_{ij}:\rangle _{0}=0$. The subtracted
vacuum divergence is  defined in Minkowski space-time and will not
affect the dynamics of the system.
%The general solution to the
%Langevin equation described above is given by \begin{equation}
%v(\tau )=e^{-\frac{\xi }{m}\tau }v(0)+\frac{1}{m}\int _{0}^{\tau
%}e^{-\frac{\xi }{m}(\tau -t')}(\eta (t')+F(t'))\, dt'\,
%.\end{equation}
% If the mirror is initially at rest, then the expectation value of
%the velocity is found to be\[
%\langle v(\tau )\rangle =\frac{1}{m}\int _{0}^{\tau }e^{-\frac{\xi }{m}(\tau -t')}\langle :F(t'):\rangle \, dt'\, .\]
%The expectation value of the normal ordered operator is taken with
%respect to some quantum state which we will specify later. Here we
%can see that the dynamics of the mirror.
Now we want to investigate the dynamics of  fluctuations of
quantum radiation pressure by calculating the velocity
fluctuations of the mirror. From the general argument related to
the fluctuation-dissipation theorem, one may  expect dissipation
occurs with respect to the fluctuating radiation force even in
vacuum \cite{FV,Polevoi}.  However, it turns out that the
associated dissipative force arising from the creation of quantum
radiation by a moving mirror has no significant effects under
normal circumstances, and thus can be ignored as compared with the
dissipative force ( $\xi$-term in Eq.(\ref{langevin}) ) from the
thermal noise that we will discuss later. We then further assume
that the quantum fluctuations of incident radiation and the
thermal noise of the fluid are uncorrelated since they are from
different sources. Thus,  the dispersion of the velocity can be
written as the sum of contributions from thermal noise and quantum
radiation pressure with respect to some quantum state to be
specified later respectively:
\begin{equation} \langle \Delta v^{2}\rangle =\langle v^{2}\rangle
-\langle v\rangle ^{2}=\langle \Delta v^{2}\rangle _{T}+\langle
\Delta v^{2}\rangle _{\rm{RP}}\, ,\end{equation}
 where \begin{equation}
\langle \Delta v^{2}(\tau )\rangle _{\rm{T}}=\frac{1}{m^{2}}\int
_{0}^{\tau }\int _{0}^{\tau }e^{-\frac{\xi }{m}(\tau
-t_{1})}e^{-\frac{\xi }{m}(\tau -t_{2})}\, \langle \eta
(t_{1})\eta (t_{2}))\rangle \,
dt_{1}dt_{2}\label{eq:dv-eta}\end{equation}
 and \begin{equation}
\langle \Delta v^{2}(\tau )\rangle _{\rm{RP}}=\frac{1}{m^{2}}\int
_{0}^{\tau }\int _{0}^{\tau }e^{-\frac{\xi }{m}(\tau
-t_{1})}e^{-\frac{\xi }{m}(\tau -t_{2})} \, \left(\langle F(t_{1})
\, F(t_{2}) \rangle \,  - \, \langle F(t_{1})\rangle \, \langle
F(t_{2})\rangle \right)\, dt_{1}dt_{2}\, .
\label{eq:dv-F}
\end{equation}
Using the noise correlation functions in Eq.(\ref{correlation}),
the first term can lead to the known result of the form
\begin{equation}
\langle \Delta v^{2}(\tau )\rangle
_{T}=\frac{2K_{\rm{B}}T}{m}(1-e^{-\frac{2\xi \tau }{m}})\, .
\label{eq:dv-eta-1}
\end{equation}
Notice that the force-force correlation function in
Eq.(\ref{eq:dv-F}) above can be expressed as the area integral of
the correlation function of the stress tensors which is divergent
in the coincidence limit where the method of regularization has to
be introduced later to remove the divergence
consistently~\cite{Barton}.  However, it is more convenient to
write this correlation function in terms of the normal ordered
operator $:F :\,= F-\langle F \rangle _{0}$ using the identity
which can be justified straightforwardly:
\begin{equation}
\langle F(t_{1}) \, F(t_{2} \rangle -\langle F(t_{1} \rangle \,
\langle F(t_{2}\rangle  =\langle :F(t_{1}):\, :F(t_{2}):\rangle
-\langle :F(t_{1}):\rangle \langle :F(t_{2}):\rangle \, ,
\end{equation}
where $\langle \rangle _{0}$ refers to taking the expectation
value with respect to the Minkowski vacuum. Here we  would like to
emphasize that in fact, the force-force correlation function that
we try to compute here is {\it not} the fully normal-ordered one
as you can see later.
 We now follow the approach of Ref.\cite{WF} by decomposing the
 force-force correlation function into the following three terms according to
the Wick's theorem:
\begin{equation} \langle :F(t_{1}):\,
:F(t_{2}):\rangle =\langle :F(t_{1})F(t_{2}):\rangle +\langle
:F(t_{1}):\, :F(t_{2}):\rangle _{\rm{cross}}+\langle :F(t_{1}):\,
:F(t_{2}):\rangle _{0}\,.
\end{equation}
The first term is the fully normal-ordered term, while the last
term is the pure Minkowski vacuum term which can be ignored as
long as we are only interested in the difference between a given
quantum state and the vacuum state. The cross terms contain the
products of the normal-ordered two point function and the vacuum
two point function, and are divergent in the coincident limit. For
a single mode coherent state $|\alpha \rangle $, the expectation
value of the fully normal-ordered force-force correlation can be
shown to be equal to the product of the expectation value of the
forces
\begin{equation}
\langle :F(t_{1})F(t_{2}):\rangle _{\alpha }=\langle
:F(t_{1}):\rangle _{\alpha } \langle :F(t_{2}):\rangle _{\alpha
}\,.
\end{equation}
 Then, the velocity fluctuation is now totally due to the cross term
of the stress tensor two point function
\begin{equation} \langle
\Delta v^{2}\rangle _{\rm{RP}}=\frac{1}{m^{2}}\int _{0}^{\tau
}\int _{0}^{\tau }\int _{A}\int _{A}e^{-\frac{\xi }{m}(\tau
-t_{1})}e^{-\frac{\xi }{m}(\tau -t_{2})}\langle :T_{xx}(t_{1}):\,
:T_{xx}(t_{2}):\rangle _{\rm{cross}}\, da_{1}da_{2}\,
dt_{1}dt_{2}\,,
\end{equation}
 One can show that the cross term of the stress tensor two point function
depends on the $z$ component of $B$ field only
\begin{equation}
\langle :T_{xx}(t_{1},x_{1}):\, :T_{xx}(t_{2},x_{2}):\rangle
_{\rm{cross}}=\langle
:B_{z}(t_{1},x_{1})B_{z}(t_{2},x_{2}):\rangle _{\alpha }\, \langle
B_{z}(t_{1},x_{1})B_{z}(t_{2},x_{2})\rangle _{0}\,,
\end{equation}
for a linearly polarized plane wave normally incident to a mirror
in the $x$ direction and with the polarization vector chosen to be
in the $y$-direction. We then decompose the $z$-component of the
magnetic field into the mode functions as
\begin{equation} B_{z}(x)=\sum _{\omega
}\sqrt{\frac{2\omega }{V}}(a_{\omega }\, \cos (\omega x)\,
e^{-i\omega t}+a_{\omega }^{\dagger }\, \cos (\omega x)\,
e^{i\omega t})\, ,
\end{equation}
 where $a$ and $a^{+}$ are the annihilation and creation operators
of the quantum field for a box normalization in a volume $V$. The
coherent state is an eigenstate of the annihilation operator
\begin{equation}
a_{\omega '}|\alpha \rangle =\delta _{\omega '\omega }\alpha |\alpha \rangle \, ,
\end{equation}
 where $\alpha $ is a complex number. Then, the velocity fluctuation
becomes (for details, see Ref.\cite{WF})
\begin{equation}
\langle \Delta v^{2}\rangle
_{\rm{RP}}=\frac{16\, \omega \, |\alpha |^{2}}{\pi ^{2}m^{2}\,
V}\int _{A}\int _{A}\, J\, da_{1}da_{2}\,
,\label{eq:dv-f-j}
\end{equation}
 where
\begin{equation}
J=\int _{0}^{\tau }\int _{0}^{\tau }e^{-\frac{\xi }{m}(\tau -t_{1})}
e^{-\frac{\xi }{m}(\tau -t_{2})}\frac{(t_{1}-t_{2})^{2}-a}{((t_{1}-t_{2})^{2}-b^{2})^{3}}
\cos (\omega \, t_{1})\cos (\omega \, t_{2})dt_{1}dt_{2}\, ,
\end{equation}
\begin{equation}
a=(z_{1}-z_{2})^{2}-(y_{1}-y_{2})^{2}
\end{equation}
and
\begin{equation} b^{2}=(z_{1}-z_{2})^{2}+(y_{1}-y_{2})^{2}\,
,\end{equation}
 where $\omega $ is the angular velocity of an incident monochromatic
laser beam. It is evident that the dissipative effects  can reduce
the mirror's velocity fluctuations that arise from the quantum
radiation pressure with a typical relaxation time scale $\tau
\approx m/\xi $. Notice that $ \langle \Delta v^{2}\rangle
_{\rm{RP}} $  in Eg.(\ref{eq:dv-f-j}) exhibits  no explicit
temperature dependence since  radiation pressure driven by a laser
beam on the mirror that we consider here in order to apply our
study to the laser interferometer comes from the quantum
mechanically coherent state. Thus, the temperature effects on the
mirror's velocity fluctuations are only implicitly though the
friction coefficient $\xi$.

Now it is of interest to study how the effects of dissipation
could possibly reduce the mirror's position uncertainly that can
be applied to the reduction of quantum noise in laser
interferometer. To so do, let us consider the time $\tau $ where
$\tau \ll m/\xi $ during which the velocity fluctuations from the
thermal noise have not grown to become significantly large, but
the time scale $\tau $ being still much larger than the intrinsic
scale of the quantum source of radiation pressure, i.e., $\tau \,
\omega \gg 1$. Taking the series expansion of the integral in
terms of the small dimensionless parameter $\xi \tau /m$ leads to
the following simple result, that is,
\begin{equation}
\langle \Delta v^{2}\rangle _{\rm{RP}}\cong \, \langle \Delta
v^{2}\rangle_{\rm{RP}}^{0}\, \, (1-\frac{\xi }{m}\tau )\, ,
\end{equation}
 where $\langle \Delta v^{2}\rangle _{\rm{RP}}^{0}$ refers to
the mirror's velocity fluctuations from the quantum radiation
pressure in vacuum which has been extensively studied in
\cite{WF}, and is of the form
\begin{equation}
\langle \Delta v^{2}\rangle _{\rm{RP}}^{0}\cong \, 4\, \frac{A\,
\omega \, \rho }{m^{2}}\, \tau \, . \label{eq:v^2_vacuum}
\end{equation}
 $\rho $ is the mean energy density of an incident laser beam that
can be linked to the laser power $P$ by $P=A\rho $ where $A$ is
the cross section area of the incident laser beam. Thus, the root
mean squared position uncertainly of the mirror is given by
\begin{equation}
\Delta \, x_{\rm{RP}}\cong \, \frac{\sqrt{wP}}{m}\tau
^{\frac{3}{2}}\, (1-\frac{\xi }{2m}\tau )\, .
\end{equation}
On the other hand, there is another source of quantum noise in the
laser interferometer, namely photon counting error, that result in
the uncertainty of the location of the interference fringe of
order~\cite{Caves1, Caves2}
\begin{equation}
\Delta \, x_{\rm{PC}}\cong \frac{1}{2\, \sqrt{\omega P\tau }}\,
.\end{equation}
 Now we would like to recall that in fact a key limit to the sensitivity
of such position measurement comes from the Heisenberg uncertainty
principle. The reduction of the uncertainty of the position
measurement is accompanied by an increase in the uncertainty in
the momentum measurement. The limit to sensitivity is referred to
as the ''standard quantum limit'', that is, in a measurement of
duration $\tau $, the minimum possible uncertainty in the
determination of the position given by~\cite{Caves1,Caves2}
\begin{equation}
(\Delta x)_{\rm{SQL}}=\sqrt{\frac{\tau }{m}} \, .
\end{equation}
 This is an intrinsic quantum uncertainty where the error of any position
measurement can hardly lies below this quantum limit. Various
works have been devoted to the use of constructive states, or
squeezed light to reduce the quantum noise and therefore increase
the sensitivity of the measurement even further \cite{WGU,WTN}.
Here we may expect that the dissipative effects from the viscous
fluid will also reduce the quantum noise on the mirror's position
uncertainly. To know whether or not it is true, we now minimize
the sum of these squared position uncertainties with respect to
the power $P$ to find a minimum uncertainty of the mirror's
position~\cite{Caves1,Caves2}. By doing so, the optimum laser
power is found to be
\begin{equation} P_{\rm{opt}}\cong \, \frac{m}{2\omega \tau
^{2}}\, (1+\frac{\xi }{2m}\tau )\, ,\end{equation}
 with the position uncertainty \begin{equation}
\Delta \, x_{\rm{Q}}\cong \, \sqrt{\frac{\tau }{m}}\, (1-\frac{\xi
}{4m}\tau )\, <\, \Delta x_{\rm{SQL}}.\end{equation}
 Presumably, one may conclude that the quantum noise on the mirror's
position measurement can be reduced to even below the standard
quantum limit due to the dissipative effects from the interaction
with the viscous fluid. However, according to the
fluctuation-dissipation theorem, the associated thermal noise with
respect to the dissipative force will cause further mirror's
position fluctuations by the amount which can be estimated from
Eq.(\ref{eq:dv-eta-1}) in the short time limit $\xi \tau/m \ll 1$:
\begin{equation}
\Delta \, x_{\rm{T}}\cong \, \sqrt{\frac{K_{\rm{B}}T}{2m}}\, \tau
\, \left(\frac{\xi \tau }{m}\right)^{\frac{1}{2}}\, .
\end{equation}
 Then, we find that the net position uncertainty, that is the root
mean square of the uncertainties from both quantum and thermal
noises, becomes
\begin{equation} \Delta \, x_{\rm{net}}\cong \,
\sqrt{\frac{\tau }{m}}\, \left[1+\left(\frac{K_{\rm{B}}T}{\hbar
\tau ^{-1}}-1\right)\frac{\xi }{4m}\tau \right]\, .
\end{equation}
 Admittedly, the net position uncertainty is larger than the standard
quantum limit for $K_{\rm{B}}T\gg \hbar \tau ^{-1}$ that is
consistent with the short time limit as well as the typical energy
scales of the thermal and quantum noises respectively. The above
calculation in fact illustrates the fact that the
fluctuation-dissipation theorem, that is to state that the thermal
noise is accompanied with sources of dissipation~\cite{PRS,DSL},
indeed plays an essential role in determining the fluctuations of
the system. Thus, this result provides a simple example to
illustrate the fact that any macroscopic treatments in order to
systematically reduce sources of noise and achieve the maximum
sensitivity of the measurement in the laser interferometer must be
consistently considered by including both fluctuation  and
dissipation effects that are correlated following the
corresponding fluctuation-dissipation theorem~\cite{JL}.

The behavior of the mirror in the late time limit, $\xi \tau /m\gg
1$ is also of interest to us from which one can study the dynamics
of fluctuations of quantum radiation pressure beyond the
relaxation time.  We first carry out the spatial integral over the
area of the mirror exposed by a laser beam in Eq.(\ref{eq:dv-f-j})
by assuming that the incident photon flux is uniform over this
area, $A=\pi \, R^{2}$. We also assume that $\omega \, R\gg 1$ to
simplify the calculations \cite{WF}. We then change the
integrating variables $u=t_{1}-t_{2}$ and $v=t_{1}+t_{2}$, and
Eq.(\ref{eq:dv-f-j}) can be written as
\begin{equation} \langle \Delta v^{2}\rangle
_{\rm{RP}}=\frac{16\, \omega \, |\alpha |^{2}}{\pi ^{2}m^{2}\,
V}\,I\, ,\label{eq:dv^2}\end{equation}
 where
 \begin{equation}
I=\frac{\pi ^{2}R^{2}}{8}\left(\int _{-\tau }^{0}du\, \int _{-u}^{u+2\tau }dv+\int _{0}^{\tau }du\, \int _{u}^{2\tau
-u}dv\right)\, \left(e^{-\frac{\xi }{m}v}\, u^{2}\, \left(\frac{1}{(u^{2}-R^{2})^{2}}-\frac{1}{u^{4}}\right)\, \left(\cos
(\omega (v-2\tau ))+\cos (\omega \, u)\right)\right)\, .\end{equation}
 We do $v$-integral and write the result in terms of the dimensionless
variable $x=\xi \, u/m$ as
\begin{eqnarray}
I & = & \frac{\pi ^{2}\, R^{2}}{4\, (\xi ^{2}+m^{2}\omega ^{2})}\, \int _{0}^{\frac{\xi \tau }{m}}\, dx\, e^{-x}\, x^{2}\, \left(\frac{1}{(x^{2}-(\frac{\xi \, R}{m})^{2})^{2}}-\frac{1}{x^{4}}\right)\, \left(f_{1}(\frac{m\, x}{\xi })-f_{2}(\frac{m\, x}{\xi })\right)\, \nonumber \\
 & = & I_{1}+I_{2}\, ,\label{eq-I}
\end{eqnarray}
 where
 \begin{equation}
f_{1}(\frac{m\, x}{\xi })=(\xi ^{2}+m^{2}\omega ^{2})\cos (\frac{m\, \omega \, x}{\xi })+\xi ^{2}\cos (\frac{m\,
\omega \, x}{\xi }-2\, \omega \, \tau )-m\, \omega \, \xi \, \sin (\frac{m\, \omega \, x}{\xi }-2\, \omega \,
\tau )
\end{equation}
 and
 \begin{equation}
f_{2}(\frac{m\, x}{\xi })=e^{2\, (x-\frac{\xi \tau }{m})}\left((2\, \xi ^{2}+m^{2}\omega ^{2})
\cos (\frac{m\, \omega \, x}{\xi })+m\, \omega \, \xi \, \sin (\frac{m\, \omega \, x}{\xi })\right)\,
 .\label{eq:it-2}
 \end{equation}
 Notice that since this integrand exhibits the singular behavior at
$x=\xi \, R/m$, one can define the above integrals in terms of the
principal value. In addition, these integrals suffer from the
intrinsic ultraviolet divergence from small  $x$ originally coming
from the coincidence limit, and the method of integrations by
parts will be implemented later to remove these divergences
consistently. Now we restrict our attention to the case of the
limit, $\xi \, R/m\ll 1$ which is relevant to the  ground-based
laser interferometer gravitational wave detector as we will study
later. Then the above integrals can be further simplified by
expanding the term in the integrand of $I$ in terms of small
parameter $\xi \, R/m$ as
\begin{equation} \frac{1}{(x^{2}-(\frac{\xi \,
R}{m})^{2})^{2}}-\frac{1}{x^{4}}\cong \frac{2}{x^{6}}\,
\left(\frac{\xi \, R}{m}\right)^{2}+{O}\left(\frac{\xi \,
R}{m}\right)^{3}\, \cdot \cdot \cdot \,
,\label{eq:expand}\end{equation}
 and keeping the most dominant term only.  Then the singular behavior at
$x=\xi \, R/m$  can be removed under this approximation.  Here we
choose to use the following  integration by parts prescription to
deal with the intrinsic ultraviolet divergence for  small $x$
typically of form:
\begin{equation} \int _{0}^{\infty }\, dx\, \frac{g(x)}{x^{4}}\,
,\end{equation}
 where $g(x)$ is a well-behaved function. The basic idea is to replace
$\, 1/x^{4}\, $ with $\, -(1/12)\, \partial _{x}^{4}\, \ln
[x^{2}]\, $, then do the integration by parts, and drop the
divergent surface terms. It can be illustrated as follows:
\begin{equation} \int \, \frac{g(x)}{x^{4}}\, dx=\frac{-1}{12}\int
\, g(x)\,
\partial _{x}^{4}\ln [x^{2}]\, dx=\frac{-1}{12}\int \, \partial
_{x}^{4}g(x)\, \ln [x^{2}]\, dx\, .\end{equation} The surface
terms can be removed by assuming  that the laser beam is  switched
on and off adiabatically in time \cite{WF}. This regularization
method has also been employed by various authors under the name of
{}``generalized principal value integration''\cite{Davies} or
{}``differential regularization''\cite{FJL}. In the following, we
will analyze the integrals $I_{1}$ and $I_{2}$ in Eq.(\ref{eq-I})
separately.

As for the integral $I_{1}$, we can rewrite this integral by using
$\int _{0}^{\xi \tau /m}\, du= (\int _{0}^{\infty }-\int _{\xi \,
\tau /m}^{\infty }) \, du\, $ as well as the expansion in
Eq.(\ref{eq:expand}) as
\begin{eqnarray}
I_{1} & \cong & \frac{\pi ^{2}\, R^{2}}{2\, (\xi ^{2}+m^{2}\omega
^{2})}\, \left(\frac{\xi \, R}{m}\right)^{2}\, \left(\int
_{0}^{\infty }\, -\int _{\frac{\xi \tau }{m}}^{\infty }\,
\right)\, dx \, \left( \frac{e^{-x}}{x^{4}}\, f_{1}(\frac{m\, x}{\xi }) \right) \nonumber \\
& =& I_{11} + I_{12} \, .
 \label{eq:I-1-appro}
 \end{eqnarray}
 We now apply the method of integration by parts to the first integral
in Eq.(\ref{eq:I-1-appro}) which exhibits the ultraviolet
divergence  for small $x$ and then carry out the integral
straightforwardly given by \begin{eqnarray}
I_{11} & = & \frac{\pi ^{2}\, R^{2}\, m^{3}\omega ^{3}}{24\, \xi ^{2}}\, \left(\frac{\xi \, R}{m}\right)^{2}\, \left(2\, \arctan (\frac{m\, \omega }{\xi })\left(1-\frac{3\, \xi ^{2}}{m^{2}\omega ^{2}}-\frac{2\, \xi ^{2}}{m^{2}\omega ^{2}}\cos (2\omega \tau )+(\frac{\xi }{m\omega }-\frac{\xi ^{3}}{m^{3}\omega ^{3}})\sin (2\omega \tau )\right)\right.\nonumber \\
 &  & \quad \quad \quad \quad \left.+\left(\gamma +\ln (1+\frac{m^{2}\omega ^{2}}{\xi ^{2}})\right)\left(\frac{\xi ^{3}}{m^{3}\omega ^{3}}-\frac{3\, \xi }{m\omega }-(\frac{\xi }{m\omega }-\frac{\xi ^{3}}{m^{3}\omega ^{3}})\cos (2\omega \, \tau )-\frac{2\, \xi ^{2}}{m^{2}\omega ^{2}}\, \sin (2\omega \tau )\right)\right)\nonumber \\
 & \cong  & \frac{\pi ^{3}R^{4}m\, \omega ^{3}}{24\xi }\, \left(1+O\left(\frac{\ln \frac{m\, \omega }{\xi }}{\frac{m\, \omega }{\xi }}\right)\right)\, ,\label{eq:I-b1}
\end{eqnarray}
 where $\gamma =0.5772157\cdot \cdot \cdot $ is Euler's constant.
In the last step, we have used the fact that $m\, \omega \gg \xi $
to be consistent with two previous assumptions, namely $\omega \,
R\gg 1$ and $\xi \, R/m\ll 1$ and the result of the integral can
be further approximated by keeping the dominant term only in this
limit . However, the second integral in Eq.(\ref{eq:I-1-appro}) is
expected to give the exponential decay behavior in the long time
limit $\xi \tau /m\gg 1$ which is given by
\begin{eqnarray}
 I_{12} & \cong  & \frac{\pi ^{2}\, R^{2}\, m^{3}\, \omega ^{3}\, \xi }{2\, (\xi ^{2}+m^{2}\omega ^{2})^{2}}\, \left(\frac{\xi \, R}{m}\right)^{2}\, \left(\frac{m}{\xi \, \tau }\right)^{4}\, \, e^{-\frac{\xi \, \tau }{m}}\, \left(\left(1-\frac{\xi ^{2}}{m^{2}\, \omega ^{2}}\right)\sin (\omega \tau )-\frac{2\, \xi ^{3}}{m^{3}\, \omega ^{3}}\, \cos (\omega \tau )\right)\nonumber \\
 & \cong  & \frac{\pi ^{2}R^{4}m\, \omega ^{3}}{2\, \xi }\left(\frac{1}{\omega ^{4}\, \tau ^{4}}\, \right)\, e^{-\frac{\xi \, \tau }{m}}\, \left(\left(1-\frac{3\, \xi ^{2}}{m^{2}\, \omega ^{2}}\right)\, \sin (\omega \, \tau )+O\left(\frac{\xi }{m\, \omega }\right)^{3}\right).\label{eq:I-b2}
\end{eqnarray}
 Substituting the results of Eq.(\ref{eq:I-b1}) and Eq.(\ref{eq:I-b2})
into Eq.(\ref{eq:I-1-appro}), we obtain \begin{equation}
I_{1}\cong \frac{\pi ^{3}R^{4}m\, \omega ^{3}}{24\xi
}\left(1+e^{-\frac{\xi \, \tau }{m}}\, \frac{12\, \sin (\omega \,
\tau )}{\pi \, \omega ^{4}\, \tau ^{4}}\, \left(1-\frac{3\, \xi
^{2}}{m^{2}\omega ^{2}}\right)\right)\, .
\label{eq:I1-result}\end{equation}
 We then follow the similar approach as the calculation of the integral
$I_{1}$ to compute the integral $I_{2}$ in the large time limit
leading to \begin{eqnarray} I_{2}
 & \cong  & \frac{-\pi ^{2}\, R^{2}\, m^{3}\, \omega ^{3}\, \xi }{2\, (\xi ^{2}+m^{2}\omega ^{2})^{2}}\, \left(\frac{\xi \, R}{m}\right)^{2}\, \left(\frac{m}{\xi \, \tau }\right)^{4}\, \left(\left(1+\frac{3\, \xi ^{2}}{m^{2}\omega ^{2}}\right)\sin (\omega \tau )+2\frac{2\, \xi ^{3}}{m^{3}\omega ^{3}}\, \left(\cos (\omega \, \tau )-e^{-\frac{\xi }{m}\tau }\right)\right)\nonumber \\
 & \cong  & \frac{-\pi ^{2}R^{4}m\, \omega ^{3}}{2\, \xi }\left(\frac{1}{\omega ^{4}\, \tau ^{4}}\, \right)\, e^{-\frac{\xi \, \tau }{m}}\, \left(\left(1+\frac{\, \xi ^{2}}{m^{2}\, \omega ^{2}}\right)\, \sin (\omega \, \tau )+O\left(\frac{\xi }{m\, \omega }\right)^{3}\right)\, .\label{eq:I2-result}
\end{eqnarray}
 We plug all results of $I_{1}$ and $I_{2}$ in Eq.(\ref{eq:I1-result})
and Eq.(\ref{eq:I2-result}) respectively into Eq.(\ref{eq:dv^2}),
and finally can obtain the velocity fluctuations of the mirror
driven by quantum radiation pressure in the presence of a
dissipative environment as follows:
\begin{equation}
\langle \Delta v^{2}\rangle _{{\rm{RP}}}=\frac{32C^{2}|\alpha
|^{2}}{\pi ^{2}m^{2}}\, I=\frac{ 2\, \pi \,\rho\, \omega ^{3} \,
R^{4}}{3\, m\, \xi }\left(1-e^{-\frac{\xi \, \tau
}{m}}\left(\frac{\xi }{m\, \omega }\right)^{2}\, \frac{24\, \sin
(\omega \, \tau )}{\pi \, \omega ^{4}\, \tau ^{4}}\right)\, ,
\label{v^2}
\end{equation} where $\rho = \omega \,|\alpha |^2 / V$.
 This is one of the main results of this work. Note that this result
bears analogy to that of the classical Brownian motion with the
relaxation time scale mainly determined by the macroscopic
parameters, i.e., the ratio of the mass of the mirror to the
friction coefficient of the mirror in a fluid. Beyond the
relaxation time scale, the velocity fluctuations start to relax to
the saturated value. Not too surprisingly, this result is
strikingly contrary to the case in which the mirror's velocity
fluctuations driven by quantum radiation pressure in vacuum grows
linearly in time in Eq.(\ref{eq:v^2_vacuum}).

As for an application, one can use the result of Eq.(\ref{v^2}) to
estimate the effects of dissipation on the quantum fluctuations of
radiation pressure in the case of the laser interferometer, such
as the LIGO interferometer. Now the role of the dissipative force
is played by the mechanism of air damping due to the incessant
collisions of air molecules with the mirror of a laser
interferometer \cite {Caves1,Caves2,AA}. The mirror in the
ground-based interferometer is of mass $m\approx 10\, {\rm{Kg}}$.
The laser power is about $ P \approx 60 \, \rm{W}$ with a typical
angular velocity $\omega \approx 4 \times 10^{15}\, \rm{rad/s}$ as
well as the spot size $A= \pi \, R^2 \approx 3\times 10^{-5}\,
{\rm{m}}^{2}$. The gravitational wave detection using the laser
interferometer is performed at quite low pressure, say about $
10^{-6} \, \rm{Torr} $ where the friction coefficient of the
mirror in the air can be obtained by summing the momentum transfer
between the mirror and each of the molecules of the air. In
\cite{PRS},  the friction coefficient for the  $1 \, \rm Hz$
pendulum of mass $10 \, \rm Kg $ at pressure below $ 10^{-6} \,
\rm {Torr}$ can be estimated as $\xi \approx \, 10^{-8} \,
\rm{N}-\rm{s} /m$. In this case, one can estimate typical energy
dissipation of a mirror per oscillation due to quantum radiation
using Eq.(4.18) in Ref.[7]  which is more than twenty orders of
magnitude smaller than that of thermal noise with the numerical
value of the fiction coefficient $\xi$ above. This serves as a
consistency check on the generalized Langivin equations in
Eq.(\ref{langevin}) that we start with. Then, with all numerical
values of these parameters, the two main assumptions, i.e.,
$\omega \, R\gg 1$ and $\xi \, R/m\ll 1$ with which to obtain
Eq.(\ref{v^2}) are shown to be satisfied. The relaxation time
scale for the mirror's velocity fluctuations can be estimated from
$\tau _{\rm{relax}}\cong m/\xi \approx  10^{9}\, \rm{s}$ which is
much larger than the typical measuring time scale for a
ground-based interferometer, $\tau _{o}\approx 10^{-2}\, \rm{s}$.
In addition, in order to understand quantitatively how the air
damping effects reduce the mirror's velocity fluctuations, one can
define $\gamma$ as the ratio of the velocity fluctuations in the
air (Eq.(\ref{v^2})) to that in vacuum (Eq.(\ref{eq:v^2_vacuum})).
 Using the above numerical parameters,  we can estimate the time scale
  when the air damping effects  reduce the velocity fluctuations by
the amount, say $10\%$, i.e., $\gamma =0.1$, which is about $\tau
\approx  10^{9}\, \tau _{\rm{relax}}\approx  10^{18}\, \rm{s}$
twenty orders of magnitude larger than the measuring time scale
$\tau _{0}$. We then conclude that the air damping effects cannot
significantly reduce the mirror's velocity fluctuations from the
quantum radiation pressure within the measuring  time scale in the
case of the ground-based laster interferometer. On the other hand,
according to the fluctuation-dissipation theorem, the associated
thermal noise with respect to the air damping effects will cause
further fluctuations on the mirror's velocity in
Eq.(\ref{eq:dv-eta-1}),  which have to be reduced in addition to
the quantum noise from the radiation pressure in order to increase
sensitivity of the measurement in the laser interferometer.

In summary, quantum fluctuations of electromagnetic radiation
pressure in a dissipative environment have been studied. We
consider a perfectly reflecting mirror which is exposed to the
electromagnetic radiation pressure in a dissipative fluid at
finite temperature. The dynamics of the mirror in a fluid of
temperature $T$ can be phenomenologically described by the
Langevin equations that involve a linear friction force term  as
well as an uncorrelated (white) noise that satisfy the classical
limit of the fluctuation-dissipation theorem.  Here we use the
generalized Langevin equations by adding a fluctuating quantum
force term into the equations  to incorporate the effects of
radiation pressure using the stress tensor approach. We assume
that quantum noise from the fluctuations of incident radiation and
thermal noise of the fluid are uncorrelated since they are from
different sources. Then, we compute the velocity fluctuations of
the mirror which can be expressed as the sum of contributions from
thermal noise and quantum radiation pressure respectively. The
velocity fluctuations from thermal noise is obtained as the result
of the classical Brownian motion. However, the calculations of the
velocity fluctuations from radiation pressure involve the
expectation value of stress tensor two point function which
contains some state-dependent divergence in the coincidence limit.
The integration by parts prescription by assuming a
switch-on-switch-off procedure in laser beam is implemented to
remove the ultraviolet divergence consistently. For a linearly
polarized quantum coherent state normally incident to a mirror,
the velocity fluctuations of the mirror are obtained analytically
for both small time and large time limits. In the small time limit
before the dynamics of the mirror's velocity fluctuations is
dominated by thermal fluctuations, we find that although the
dissipative effects from the interactions of the mirror with the
molecules of the fluid can reduce quantum noise on the mirror's
position measurement below the standard quantum limit, according
to the fluctuation-dissipation theorem, the corresponding thermal
noise with respect to the dissipative force will cause further
mirror's position fluctuations that gives rise to the net position
uncertainty, that is the root mean square of the uncertainties
from both quantum and thermal noises, larger than the standard
quantum limit. Thus, this result provides a simple example to
illustrate the fact that any macroscopic treatments in order to
systematically reduce sources of noise and achieve the maximum
sensitivity of the measurement in the laser interferometer must be
consistently considered by including both fluctuation  and
dissipation effects that are correlated following the
corresponding fluctuation-dissipation theorem. The result of the
large time behavior of the velocity fluctuations of a mirror in a
dissipative environment is applied to the laser interferometer of
the ground-based gravitational wave detector. The role of the
dissipative force in this case is played by the air damping due to
the incessant collisions of air molecules with the mirror. Our
results reveal  that in fact during the measuring time scale, the
air damping effects cannot significantly reduce the mirror's
velocity fluctuations from the quantum radiation pressure. This is
our first attempt to tackle the issue of fluctuations of quantum
radiation pressure in the presence of a dissipative environment.
We note that this generalized  Langevin  equations using the
stress tensor approach can be applied to more complicated case,
for example, the dynamics of the mirror that undergoes the
pendulum motion involving thermal noise of its suspension, while
it is exposed by the laser beam\cite{AA}. The results of this work
will have direct impact on the laser interferometer gravitational
wave detector, and the work is underway.

\vskip 0.5cm

We would like to thank  Daniel Boyanovsky, Hector de Vega, Larry
Ford, and Bei-Lok Hu for their useful discussions. The work of
C.-H. Wu and D.-S. Lee were supported in part by the National
Science Council, ROC under the grants NSC91-2112-M-259-008.

%\begin{thebibliography}{10}

%\end{thebibliography}

\end{document}